# Discussion about the assumptions of Category Theory approach to agent-based modeling in microeconomics

Panu Jalas

*Kerttu Saalasti Institute, University of Oulu, Pajatie 5, 85500 Nivala*

______________________________________________________________________

ABSTRACT

We investigate a possible category theoretical description for agent based modeling by outlining justifications for two main principles to describe the valuations in a realistic way in microeconomics: 1) It is assumed that the valuations can be expressed as a subcategory of the category of metric space so that value differences between various objects, that agents assign to them, can be expressed with a metric that is consistent with the concept of distance in mathematical metric spaces. 2) In realistic economic models, the category of valuations does not consist of linear spaces other than in special cases.

We also discuss how category theoretical concepts such as morphisms and functors can be used to express transformations within categories and relations between other categories. We then present examples how functors and morphisms could be used to describe relationships and operations, such as ownership changes, transactions, and price formation, in the context of some of the established theories in microeconomics. Finally, we discuss briefly possible applications, such as product design and price optimization.

______________________________________________________________________

1. INTRODUCTION

In recent years, there has been a growing amount of interest in category theory approach to provide an intuitive mathematical framework for understanding economic modeling. Specifically, in microeconomics, category theory has potential to provide ways to formulate very general level mathematical descriptions about economic interactions. The philosophical and historical background for the benefits of category theory for economic modeling has been discussed recently *e.g.* by Crespo and Tohmé (2017) and Lanzi (2022). Specific proposals for

the use of category theory in economics include, for example, a model for credit risk by Tran et al. (2021)

The intent of this article is to study how category theory could provide a realistic general-level framework for creating agent-based models in microeconomics (agent based modeling in *e.g.* Tesfatsion, 2023 or Axtell and Doyne, 2022). For that purpose, we discuss some of the key principles of category theory and outline a proposal, Category Theoretical Framework for MicroEconomic modeling (CTFME), in order to investigate how realistic and practical agent based microeconomic models could be described with category theoretical concepts. The main challenge with creating the framework was to understand and justify the universality of the mathematical principles, how valuations, created by cognitive actors acting as economic agents, can be realistically used in microeconomic modeling when the valuations are essentially immaterial entities imagined by the agents.

The motivation of this article is also to promote cross-disciplinary discussion about the category theory and mathematical foundations of microeconomic modeling. For example, in CTFME we make the assertion that usually the calculation with subjective valuations cannot incorporate assumption of mathematical linear spaces. While linearity is intrinsically assumed in models, describing natural sciences or engineering, we discuss reasons, why in our opinion linearity is usually not applicable for microeconomic models involving cognitive actors.

In CTFME, we discuss justifications for two main principles how to describe the valuations in realistic models: 1) It is assumed that the valuations can be expressed as a subcategory of the category of metric spaces. So when modeling microeconomics, we can assume that the subjective opinions, which agents have about the value difference between various objects, can be expressed with a metric that is consistent with the concept of distance in mathematical metric spaces. 2) In realistic economic models, the category of valuations does not consist of linear spaces other than in special cases. Nonlinearity means that the microeconomic models within CTFME can result in a very different calculus compared to models in physical space-time world and engineering, where a linear vector space is always assumed as the foundational framework for all modeling.

In Chapter 2, we outline the CTFME framework for agent-based modeling by assigning mathematical categories to represent objects, agents, the properties of objects and the valuations that the agents about the properties. In Chapter 3, we discuss how to justify the two key assumptions of CTFME: The category of valuations is 1) a subcategory of the category of metric spaces, but 2) usually it is not a linear space. In Chapter 4, we discuss how category theory concepts can be used to express relations between members within categories and those within other categories. We also make some proposals how functors could be used to describe relationships and operations, such as ownership, transactions, and price formation, in the context of some of the established theories in microeconomics, such as choice modeling (*e.g.,* McFadden 1980, Anderson, 1992) or prospect theory (Kahneman and Tversky, 1984). In Chapter 5, we briefly discuss applications, such as product design and price optimization. Finally, in Chapter 6, we present conclusions and summarize ideas for further study.

## 2. CATEGORY THEORETICAL FRAMEWORK FOR AGENT-BASED MODELING IN MICROECONOMICS

In the following, we define CTFME framework by adapting the most basic concepts and terminology of established agent-based modeling theories and by creating representative expressions with mathematical category theory notation following *e.g.,* Marquis, J.-P. (2015):

A1 *Objects and properties*

We call the entities that are subject to economic transactions objects $O_i$. We assume that each $O_i$ can be represented by the set of its properties $p_i = \{q_1, q_2... q_n\}$ so that $p_i$ defines objects sufficiently such that if two objects $O_i$ and $O_j$ can be characterized by the same set $p_i$, they can be considered identical from any agent's point of view. For example, if two products have the same set of product specifications, from a buyer's perspective, they are similar; thus, their value for any buyer is the same.

A2 *Agents and valuations*

We call people, organizations or other legal entities, which can own objects, as agents $A_j$. We assume that the concept of valuation is the representation of the desirability or value that an agent associates with an object. We then denote the valuations that agent $A_j$ has about object $O_i$ and its property set $p_i$ with a set $v_{ij} = \{w_{1j}, w_{2j}... w_{mj}\}$, where each $w_{kj}$ is the valuation that agent $A_j$ has with respect to the corresponding property $q_k$ of $p_i$.

For a general-level framework, we do not assume any other structure for valuations that relate to motivations, such as utility, desirability, or need, which were the underlying causes for the agents to assign different valuations to different properties. However, incorporating specific motivational factors (*e.g.,* Axtell and Doyne, 2022, chapter III) can be represented in individual models based on the framework.

We define the CTFME framework by postulating that the sets of properties of objects and the valuations that agents have about the property can be represented with the following categories:

C1. The category of the sets of the properties of the objects is **Pob**
    a. **Pob** consists of sets $p_i$, which define objects $O_i$ such that $O_k$ and $O_l$ are indistinguishable for any agent if $p_k = p_l$.
    b. **Pob** is a subcategory of **Set**, the category of all sets.

C2. The category of the sets of valuations is **Val**
    a. **Val** consists of sets of valuations $v_{ij}$ that agents $A_j$ relates to objects $O_i$ and their property sets $p_i$
    b. **Val** is a subspace of **Met**, the category of all metric spaces

From a practical point of view, postulate C1 does not impose many restrictions on the nature of the objects other than that they are defined by their properties such that the set of properties provides all the relevant information for the agent to form an opinion about the value of the object.

Postulate C2 assumes that the sets of valuations that the agent can have about the objects have more structure than the sets of properties. The key feature of a metric space **Met** is the distance function. We posit that for a microeconomic model, connecting preferences or other opinions to an actualized market price, described with a real number, assuming the existence of a metric is useful and intuitively even necessary. However, we suggest that assigning a more restricted mathematical space, with the assumption of linearity for the postulation of **Val**, does not usually yield a realistic framework for microeconomic modeling. In contrast, we argue that in most cases, the valuation sets are not linear spaces, and consequently, the calculus used in agent-based models must account for that nonlinearity.

## 3. VALUE DIFFERENCE AS A METRIC AND NON-LINEARITY

To justify the postulation of **Val** as a subspace of **Met,** a practical interpretation of members of **Val**, representing valuations between the property sets of objects, should exhibit the notion of distance $d$ between each other so that $d$ can be expressed as a real number. It is related to the perceived value difference that an agent assigns between two objects $O_i$. However, for **Val** to be a metric space, its members $v_{ij}$ must satisfy the following conditions with respect to the metric $d$:

Let $x$, $y$ and $z$ denote the valuation sets $v_{xj}$, $v_{yj}$, and $v_{zj}$, respectively, which are members of **Val**. For **Val** to be a metric space, $x$, $y$ and $z$ need to comply with the axioms M1-M4. of a metric space

M1. Identity: The distance from a point to itself is zero:

$$d(x,x) = 0$$

M2. Positivity: The distance between two distinct points $x \neq y$ is always positive

$$d(x,y) > 0$$

M3. Symmetry: The distance from $x$ to $y$ is always the same as the distance from $y$ to $x$:

$$d(x,y) = d(x,y)$$

M4. The triangle inequality: Distance from one point to another is always shorter or as short as taking a detour through a third point.

$$d(x,y) \leq d(x,y) + d(x,y)$$

Conditions 1-3 can be considered relatively straightforward to justify, but Condition 4, the triangle inequality, is more complicated. We provide the following argumentation to support the hypothesis that for realistic modeling in microeconomics, it is reasonable to assume that the **Val** is a metric space:

M1.  Identity: The distance from a point to itself is zero. In the case of two identical objects, based on the postulate C1 a., it is reasonable to assume that the agent does not find any difference in their value. Hence, the value difference *d* is zero.

M2.  Positivity: When we use *d* as the metric for value difference, we suggest that when comparing two nonidentical objects, there is always at least an infinitesimal difference in perceived value when agents assign a valuation to the object. This is important, for example, when modeling a case where an agent thinks that the valuation of two nonidentical objects is very close to the same, yet it requires choosing only the other object, for example, because of being able to afford only one of the objects.

M3.  Symmetry: The assumption of symmetry is that agents assign valuations based only on an object's properties. In practice, however, there can be cases where the temporal order of agent observing the object or other environmental circumstances affect how agents assign valuations to objects. If that is the case, such circumstances should be included as the properties of the object. When objects are either presented in similar circumstances or if the description of the object's properties sufficiently (from the agent's point of view) accounts for the impacts that circumstances might have, symmetry becomes a valid assumption.

M4.  Triangle inequality: The triangle inequality indicates that when the agent compares the object with property set $p_x$ to object with properties $p_z$ and then $p_z$ to $p_y$, the difference between $p_x$ and $p_y$ cannot be smaller than that in a situation where the agent assigns $d(x,y)$ without comparisons. To justify this, we argue that the counterstatement leads to inconsistency when creating an agent based microeconomic model with valuations: If we assume that in a model there exists *z* such that $d(x,y) > d(x,z) + d(z,y)$, the model produces the value difference between *x* and *y* that increases by adding the comparison *z*, and an additional value *M* is created by making the comparison $d(x,y) - d(x,z) - d(z,y) > M > 0$. Consequently, it is always preferable for the agent to consider exchanging *x* to *z* and then *z* to *y* rather than directly exchanging *x* to *y*. However, according to M3, it is possible to exchange *y* back *x* with the same value $d(y,x) = d(x,y)$; therefore, for each cycle of exchanges, the agent ultimately creates a value difference *M* just from the exchange activity. However, this is not a realistic real world economic model because it incorporates a mechanism to create value from nothing whereas in reality exchanging objects usually costs value or, at best, can be considered free. Thus, a realistic economic model should assume that M4 is valid and adding exchanges of additional objects should be considered either negligible or produce an increase in value compared to a direct exchange of two objects.

Assuming that **Val** is a metric space, the next question is whether it is reasonable to assume that it is also a linear space.

If **Val** is a linear space, for any valuation set $v \in$ **Val**, it should be linear with respect to multiplication with a scalar such that for any scalar $\alpha, \beta \in \mathbf{R}$

L1   Associativity

$$\alpha (\beta v) = (\alpha \beta) v$$

L2   Multiplication with one

$$1 \cdot v = v$$

However, we argue that associativity according to L1 is usually not realistic in microeconomics because an agent's valuation of objects is highly dependent on the amount of objects, that the agent is assessing. For example, the total value of 100 times of bundles of 2 products is usually much more expensive than 2 bundles of 100 products. In real-life economic interactions, the valuations are dependent on the cost of bundling according to quantity. The bundling cost can depend, for example, on extra work and materials for packaging, shipping and inspection. There can be also costs related to the risks and the complexity of delivering many small orders compared to that of just a few large orders. Additionally, from the agent's point of view, the valuation is dependent on how many items of the object consider buying. It is very typical that if an agent needs just one object, it is unwilling to pay more for any additional copies. Thus, the bundling of objects, represented by the scalars $\alpha$ and $\beta$ in L1, usually impacts the agent's appreciation of the objects and the multiplication is dependent on the order of bundling which contradicts the assumption of associativity. Consequently, it can be expected that most agent-based models for microeconomics should assume a non-linear metric space as the basis for calculus yielding potentially very different outcomes compared to models assuming a linear space as the basis for scalar multiplication.

4. RELATIONS BETWEEN AGENTS AND OBJECTS

Microeconomic modeling should provide information about relations and operations between agents and objects. In the CTFME framework this can be investigated by creating relations between the categories of sets of properties and the metric spaces of valuations with the concept of a functor. A functor maps elements from one category to elements in another category and maps morphisms, which are transformations within a category (such as functions), between the elements in the first category to morphisms between corresponding elements in the second category. Functors must preserve the structure of categories, meaning they respect the composition of morphisms and identity morphisms. In earlier work, for example, Lanzi (2022) used the concept of morphism to describe how agents view objects in the context of choice modeling.

In the following, we provide the standard definitions of morphisms and functors (e.g., Marquis, 2015) and propose examples of how the concepts of microeconomics can be interpreted as morphisms and functors with category theory.

Morphisms describing transformations within a category

Morphisms represent a relationship or a mapping between two objects in a category. The term "morphism" can be interpreted as a generalization of the concept of a function or map in other branches of mathematics:

Given a category $C$, where $x$ and $y$ are elements in that category, a morphism $f$ indicates a relationship or transformation from the object $x$ to the object $y$. A morphism is represented by an arrow from object $x$ to object $y$ and it can be denoted as $f: x \rightarrow y$.

H1. Composition: If there are morphisms $f: x \rightarrow y$ and $g: y \rightarrow z$, then there exists a composition morphism $\circ \rightarrow g \circ f: x \rightarrow z$. Composition is associative, meaning that $(h \circ g) \circ f = h \circ (g \circ f)$ for morphisms $f$, $g$, and $h$ in the appropriate domains.

H2. Identity: For every element $x$ in the category, there exists an identity morphism $Id: X \rightarrow X$. The identity morphism acts as a neutral element with respect to composition, meaning $f \circ Id\ x = f$ and $Id\ y \circ f = f$ for any morphism $f: x \rightarrow y$.

H3. Closure under composition: The composition of two morphisms within the same category produces another morphism in that category.

Functors describing relationships between categories

When *C* and *D* are two categories, functor *F* from category *C* to category *D* consists of two components:

F1. Object component: For each element *x* in category *C*, there is an associated object *F(x)* in category *D*.

F2. Morphism component: For each morphism *f: x →y* in category *C*, there is an associated morphism *F(f): F(x) →F(y)* in category *D*.

In addition, functors must satisfy two conditions to be considered valid:

F3. Identity: For every object *x* in category *C*, the identity morphism *Id x: x →x* must be mapped to the identity morphism *Id {F(x)}: F(x) →F(x)* in category *D*.

F4. Composition: For every pair of morphisms *f: x →y* and *g: y →z* in category *C*, the composition of their images under the functor must be the same as the image of their composition: *F(g ∘f) = F(g) ∘F(f)*.

In the context of CTFME, we consider the following examples to describe the relationships between agents and objects:

Changes in properties and valuations as morphisms and value functions as functors to connect properties and valuations.

When creating a model within the CTFME that can describe relations (e.g., preferences or prices) between the categories of valuations **Val** and properties **Pob**, the categories should be linked with a functor. We can denote that as functor *V* from the category **Pob** to the category **Val**, corresponding to statement F1. We propose that, following the ideas of prospect theory (Kahneman and Tversky, 1984), *V* can also be interpreted as a value function with which agents relate subjective valuations to objects.

The value functions also reflect the morphism from a set of properties $p_1$ to a set $p_2$ within a category **Pob** to a change in valuations in **Val** (F2). A morphism in **Pob** can be, for example, a change in and attribute of the product, for example, its color or aging with respect to the time of manufacture. Then, there is a corresponding change in the valuations that agents have about the object from $v_{1j}$ to $v_{2j}$, where *j* refers to agent $A_j$. In practice, when creating a specific model within CTFME, in order to consider *V* as functor, the model should justify that F3 and F4 are also true.

Preferences as functors

Preferences, which agents assign between different objects, can also be described as functors between categories. For example, in the case of choice modeling (classical choice modeling, e.g., in McFadden, 1980), preferences can be represented as functors *P* between the categories of **Pob** and **Val**. If we represent differences between objects with morphisms *f* and *g* according to H1, the composition of the two morphisms *f* and *g*: $g \circ f$ is consistent with the assumption of transitivity between preferences in choice models.

Ownership and price formation in transactions

Within the CTFME framework, there can be multiple ways to represent ownership. One way to represent ownership is to describe it as a property $p_k$ of the object's property set. Then, the change in ownership can be denoted as a morphism *f* that changes the ownership property from $p_k$ to $q_k$. Furthermore, we can describe the change of valuations, how agents *P* and *Q* appreciate the Object's ownership property $p_k$ morphing to $q_k$, with functors from **Pob** to **Val** as *Vp(f)* and *Vq(f)*, respectively. When the difference in value between the Objects with respect to ownership for agent *V* is $d_{pk}$ and that for agent *W* is $d_{qk}$, we can analyze the condition $d_{pk} = d_{qk}$ as a potential threshold for a transaction to take place. Thus if both Agents *P* and *Q* agree on the value difference $d_{pk} = d_{qk}$ it can be interpreted to represent the exact numerical value that is a condition for the transaction to occur, which is the actualized *price* of the object. We can also consider that a transaction could occur in all cases that agent *P* is the seller and agent *S* is the buyer, when $d_{pk} \leq d_{qk}$, the buyer appreciates the object's change in ownership as much or more than does the seller.

The previous examples are meant only as suggestions to describe microeconomic interactions within the CTFME framework. We believe that much more investigation is needed to understand the impact of the boundary conditions set by H1-H3. and F1.-F4 to the applicability of CTFME based real-life microeconomic modeling.

5. APPLICATIONS

The assumptions of CTFME have implications for how to approach the modeling of many of the main applications of product and marketing design based on microeconomic analysis. Especially the notion that the category of valuations does not need to be a linear space yields that the scalar multiplication of valuations is often non-linear.

Product design

One of the main challenges of product design is to optimize the properties of a new product so that it can provide the maximum value for the intended customers. In CTFME, which corresponds to a model where the value of the product is a functor *V* from **Pob** to **Val,** the changes in the properties of the product design can be described as morphisms within the category. However, because of nonlinearity, the model must be able to also consider that the impact of changing a property is not linear in respect with the change in valuations. The assumption of an appropriate formula for the functor *V* would then need to be justified or validated with experimental data.

Focus Group Analysis

In focus group analysis, the relationships between the attributes of agents and their valuations are investigated. A category theoretical interpretation can be to create a category **Ag** to represent the attributes of agents a subcategory of **set**. Then, the functors between **Ag** and **Val** represent the relation between certain types of agents and their valuations and how the morphism of an attribute in a set of **Ag** is reflected in **Val**.

Artificial Intelligence

In the study of consumer behavior, AI algorithms are used to analyze large unstructured datasets from various sources. A category theoretical framework can provide a foundation for the algebra used in the algorithms. For example, when analyzing consumer behavior with AI, the assumption of metric space for the category **Val** sets a condition that when training the algorithm how the consumers form their valuations, the algorithm does not create outcomes where the consumers seem to create additional value difference between two products just by exchanging them with third products. Additionally, the assumption of nonlinearity makes it possible to produce realistic outcomes such that the optimal price does not depend linearly on the amount of products sold.

6. SUMMARY

The objective of this article is to advance the conversation surrounding the utilization of category theory in economic modeling. We constructed a framework, CTFME, with category theoretical interpretations for agents, objects, properties and valuations for agent-based modeling of microeconomics. The two key assumptions in CTFME are that 1) when creating realistic models for microeconomic analysis, it is reasonable to assume that the category **Val**, which represents subjective valuations that agents have about objects, is a subcategory of the category of metric spaces **met**. 2) The members of **Val** do not usually exhibit linearity; thus, subjective valuations cannot be represented as mathematically linear spaces. This aspect is a major difference in comparison to modeling in physics or engineering. Thus, a more general mathematical algebraic framework, such as category theory, is a very plausible candidate as a basis for realistic modeling in microeconomics. We hope that this article stimulates discussion on whether the assumptions of the general applicability of metric spaces and nonlinearity can be further justified or criticized.

We also provided some examples of how the main category of theoretical expressions for relations, morphisms and functors can be used to interpret some of the established microeconomic modeling techniques within the CTFME. However, these examples are presented only to demonstrate ways to assign category theoretical interpretations, and they have not yet been used, developed with rigorous argumentation or evaluated in practice. In the future, it is of interest to develop these ideas further. We also propose that Lanzi's (2022) suggestion to describe morphisms as means to describe the relational perspective on social action and historical change basically aims toward the same objective of representing change within a category, in a similar manner that we suggest describing the change of assigning valuations to properties in Chapter 4.

We think that a fundamental question is how powerful category theory can be in mathematically "harnessing" the description of cognitive actors' behavior within an economy. Does category theory provide sufficient means to create universal models that can be widely applied in economics and for all transactional situations? What are the limits of applicability or universality of a framework, such as the CTFME? Specifically, what are the limits of validity for the justifications of the two key postulates (valuations as metric spaces and nonlinearity) of CTFME? Also In light of advancing artificial intelligence technologies to understand consumer behavior and agent based modeling in general, we believe that this can be a very interesting field of study in order to better understand how category theory can provide new ways to realistically simulate agents' behavior as economic actors.